\documentclass[12pt]{iopart}
\usepackage{iopams}  
\usepackage{graphicx}
\usepackage{pdflscape}
\usepackage{cite}
\begin{document}


\title[Spectroscopic parameters for W$^{39+} ion$]
{Quasirelativistic calculation of 4s$^2$4p$^5$, 4s$^2$4p$^4$4d and 
4s4p$^6$ configuration spectroscopic parameters for the W$^{39+}$ ion}

\author{P. Bogdanovich,  R. Karpu{\v s}kien{\. e}, R. Kisielius}

\address{Institute of Theoretical Physics and Astronomy, Vilnius University, 
A. Go\v stauto 12,  LT-01108, Vilnius, Lithuania}


\begin{abstract}

The {\sl ab initio} quasirelativistic Hartree-Fock method developed specifically 
for the calculation of spectral parameters of heavy atoms and highly charged 
ions is used to derive spectral data for the 4s$^2$4p$^5$, 4s$^2$4p$^4$4d and 
4s4p$^6$ configurations of the multicharged tungsten ion W$^{39+}$. 
The relativistic effects are taken into account in the Breit-Pauli 
approximation for the quasirelativistic Hartree-Fock radial orbitals. 
The configuration interaction method is applied to include the electron 
correlation effects. Produced data are compared with existing experimental 
measurements and theoretical calculations.


\noindent{\bf PACS:\/} {31.15.ag, 32.70.Cs, 95.30.Ky}\\

\end{abstract}

\maketitle

\section{Introduction}
\label{intro}

Metallic tungsten is becoming one of the popular materials in high-temperature 
devices, including fusion reactors \cite{01,02}. The tungsten material is 
difficult to vaporize. Nevertheless, the highly-charged ions of tungsten can 
emerge in fusion plasma causing the decrease of its temperature. Thus there is 
a need to control the concentration of these ions by monitoring their spectra.
Only reliable theoretical data of the multicharged tungsten ion spectral 
properties enable to model such types of plasma providing means to its 
diagnostics, see e.g. \cite{03}. On the other hand, the theoretical energy
spectra and transition parameter calculations are helpful in identifying 
the emission lines of the multicharged tungsten ions which are observed in 
the specially designed experiments. 

This situation has increased interest from both the experimental and theoretical 
physicists in study of the tungsten ions of various ionization stages. 
An extensive compilation of the 
experimental data and their semi-empirical investigation has been reported in 
\cite{04}. A consequent review of the experimental and theoretical works is 
reported in \cite{05}. Over the last few years, several experimental works (see, 
e.g. \cite{06,07,08,09} and the references therein) on the tungsten ions are 
published. Alongside there is
a considerable interest in multicharged ions from theoreticians. Some part of
reported calculations are based on different variational methods utilizing the
solutions of Dirac-Fock relativistic equations, see \cite{10,11,12,13}.
Another part of calculations \cite{14,15} is mainly based on the relativistic 
many-body perturbation theory (RMBPT) with additional calculations in \cite{15} 
performed using a quasirelativistic Hartree-Fock with the superposition of 
configurations and the relativistic multiconfiguration Hebrew University 
Lawrence Livermore Atomic Code (HULLAC). A quasirelativistic approximation (QR) 
developed at Vilnius University was successfully applied for the study of 
tungsten ions in \cite{16,17,18}. The above works were devoted to the tungsten 
ions with an open 4d shell. 

The quasirelativistic investigation of the tungsten ions with an open 4d shell 
\cite{16,17,18} has demonstrated that our QR
\cite{19,20,21} is capable of producing reliable and highly-accurate 
spectroscopic parameters for the tungsten ions of such a high ionization
degree. This conclusion gives us confidence to consider the higher ionization 
stages and to start investigation of the tungsten ions with an open 4p shell.

In the current work we present the results for the W$^{39+}$ ion in the ground
and the lowest two excited configurations. The relevance and actuality of the 
spectroscopic data for this ion is reflected in number of papers; for us 
new theoretical transition data open a way to compare those data to our
calculation results and to assess the accuracy of determined spectroscopic 
parameters.

The experimental electron-beam-ion-trap (EBIT) measurements for the 
multicharged ions with an open 4p shell were reported in \cite{22}. There 
authors employed theoretical data from \cite{23} for the line identification.
The fully relativistic parametric potential code RELAC \cite{24} was adopted
in \cite{23} for the data production. Later on, some new lines of the 
W$^{39+}$ ion were presented in \cite{25} where the HULLAC \cite{26} code,
analogous to the RELAC one, was used in calculations. The EBIT measurements 
in \cite{27} made it possible to identify few more lines by applying the
Flexible Atomic Code (FAC) \cite{28}. Those experimental studies were summarized
in \cite{04}, and the atomic data could be found in database \cite{29}.

Very recently, several new theoretical studies were reported. In those works
data for the ion W$^{39+}$ ground and two lowest excited configuration level 
energies, the level radiative lifetimes $\tau$ and the electron transition 
parameters were presented. Their studies were using fully relativistic 
calculations with the configuration interaction (CI) approximation used
to include electron-correlation effects as implemented in the computer code
GRASP (General-purpose Relativistic Atomic Structure Package) \cite{30}. The 
theoretical energy levels of the ground 4p$^N$ and 4d$^N$ configurations of the 
tungsten ions and the radiative transition parameters among their levels were 
calculated in \cite{10}. The data for the excited configurations of the W$^{39+}$ 
were reported quite recently. The first calculation \cite{31} results were 
substantially analyzed and strongly criticized in \cite{32}. Based on the
latter, the same authors presented more extensive calculation results for the 
W$^{39+}$ in \cite{33}. The discussion about the relativistic calculation of 
the spectral parameters \cite{34} was extended in \cite{35}, where several 
systematic calculations of the W$^{39+}$ were performed using different
CI wavefunction expansions.

Our previous quasirelativistic calculations of the spectroscopic parameters 
for the tungsten ions with open 4d shell \cite{16, 17, 18} pointed out
to the high-accuracy of our transition probability values and their nice
agreement with relativistic calculation results from \cite{10} when the 
transitions among the the ground configuration levels were considered.
Unfortunately, the lack of calculated data did not allow to compare data for the 
transitions from the levels of the excited configurations. Above mentioned 
W$^{39+}$ studies \cite{33, 35} provided us with an impetus to perform a 
large-scale comparison of quasirelativistic and relativistic calculation data.
In the next section we give a short description of employed quasirelativistic
approach. In \sref{results} we compare our results with the data from other
authors. The summary and conclusions are presented in \sref{summ}.

\section{Description of quasirelativistic calculation}
\label{qr}

The quasirelativistic calculations are performed according to the description
given in our previous papers \cite{16, 17, 18} and the references therein. 
At the first stage, we solve the
QR equations for the ground configuration 4s$^2$4p$^5$, see \cite{19, 20}. 
Here we need to mention some peculiarities of our calculations. It was 
demonstrated in \cite{20} that, for the averaging of QR equations over the 
one-electron angular momenta $j$, one has to improve their accuracy at the 
spherical $r$-coordinate origin ($r \rightarrow 0$) by accounting for the 
spin-orbit interaction. For that purpose, an additional parameter $X_{np}$ has 
been introduced. The calculations for the tungsten ions with open $n$p shell 
demonstrated that  the $X_{np}$ value significantly influences the values of 
the spin-orbit interaction parameter $\eta(4p)$. In order to improve the 
accuracy of the spin-orbit interaction, we have increased the parameter $X_{np}$
value by few tens of percent compared to that in \cite{20}.

At the second step, we solve the QR equations for the 4d and 4f radial orbitals 
(RO) in a frozen-core potential.  In order to perform our calculations in a
multiconfiguration approximation, the basis of determined RO is supplemented 
with the transformed radial orbitals (TRO), which have variable parameters
\cite{21}. The TRO are determined for the radial orbitals having the principal 
quantum number $5 \leq n \leq 10$ and for all possible values of the orbital 
quantum number $l$, i.e. $l \leq 9$.

The relativistic corrections are included in the Breit-Pauli approach specially 
adopted for the quasirelativistic RO \cite{21}. We employ the same RO basis both
for the even and the odd configurations. Therefore we can avoid any inaccuracies 
of the calculated transition parameters occurring when the non-orthogonal RO are
used.

The set of the admixed configurations for the CI expansion was generated by
virtually exciting one or two electrons from the external $4l$ shell and the
internal $3l$ shell of the adjusted configurations 4s$^2$4p$^5$, 4s$^2$4p$^4$4d 
and 4s4p$^6$. Such a set enables us to include comprehensively the correlation
effects including the core-polarization effect. By applying this method and 
the generated RO basis for the ground (odd) configuration 4s$^2$4p$^5$, we can
determine the total amount $M_{\mathrm{o}} = 3172$ of the admixed configurations 
which can interact with the adjusted configuration. Further we apply the 
selection method for the admixed configurations. More details on this method 
are given in \cite{36}. We need to mention that this method has been applied in 
all our previously described calculations. In the present calculation, we select 
$S_{\mathrm{o}} = 188$ strongly-interacting configurations according to their 
weights in CI expansion, including the adjusted one. They produce total number 
of $C_{\mathrm{o}} = 116008$ configuration state functions (CSF). 
The $C_{\mathrm{o}}$ is further reduced to $R_{\mathrm{o}} = 2824$ by applying 
the CSF reduction methods described in \cite{37}.

For the excited (even) configurations 4s$^2$4p$^4$4d and 4s4p$^6$, one can
generate $M_{\mathrm{e}} = 7286$ admixed configurations. Our selection methods
help us to reduce this number to $S_{\mathrm{e}} = 949$ most important 
configurations. Furthermore, the initial number of CSF 
$C_{\mathrm{e}} = 6234470$ is reduced to $R_{\mathrm{e}} = 426816$ in the same
way as it is done for odd configurations.

After the two-step selection methods are applied, the determined sets of CSF
are employed to compute the level energies of the W$^{39+}$ ion. In our 
calculations, the main limiting factor is the number of the same $LS$ terms. 
It determines the size of the Hamiltonian matrices to be diagonalized. In our 
case of the W$^{39+}$ ion, the largest number, 71039, arises for the $^2$F term 
of the even configurations. This number is close to our present computational limit. 
All our calculations are starting from the non-relativistic $LS$-coupling.
After the level energies and their eigenfunctions are determined, these are 
adopted to calculate the spectroscopic parameters of the radiative transitions. 
From these data, the level radiative lifetimes are derived. During the 
investigation of the multicharged tungsten ions in \cite{16, 17} we have 
noticed that the radiative lifetimes $\tau$ of some metastable levels can
be affected by the M2 and E3 transitions to the ground configurations -
not only by the M1 and E2 transitions among the levels of the excited 
configurations. Therefore we also have calculated such transitions for the
W$^{39+}$ ion.

To perform our calculations, we have employed our own original computer codes 
together with the codes \cite{38,39,40} which have been specifically adapted 
for our computing needs. The code from \cite{38} was updated according to the 
methods presented in \cite{41,42}.

\section{Results and discussion}
\label{results}


\setlength{\tabcolsep}{3pt}

\noindent

\Table{\label{t1}
The level energies $E$ (in 100 cm$^{-1}$) of W$^{39+}$ and the main contributions
to their wavefunctions 9in \%) from QR calculations.
}
\br
\multicolumn{1}{l}{}&
\multicolumn{1}{c}{}&
\multicolumn{1}{c}{}&
\multicolumn{1}{c}{}&
\multicolumn{1}{c}{NIST}&
\multicolumn{1}{c}{Exp}&
\multicolumn{1}{c}{Exp}&
\multicolumn{1}{c}{QR}&
\multicolumn{1}{c}{RELAC}&
\multicolumn{1}{c}{MCDF}&
\multicolumn{1}{c}{GK$_{\mathrm{CV}}$}\\
\ns
\multicolumn{1}{l}{$N$}&
\multicolumn{1}{c}{Configuration}&
\multicolumn{1}{c}{$J$}&
\multicolumn{1}{c}{\%}&
\multicolumn{1}{c}{[29]}&
\multicolumn{1}{c}{[22]}&
\multicolumn{1}{c}{[25]}&
\multicolumn{1}{c}{}&
\multicolumn{1}{c}{[22]}&
\multicolumn{1}{c}{[33]}&
\multicolumn{1}{c}{[35]}\\
\ns
\mr                                           
1    & 4s$^2$4p$^5$          $^2$P     & 1.5 &98& 0   	&       &       & 0     & 0           & 0           & 0	    \\
2    & 4s$^2$4p$^5$          $^2$P     & 0.5 &98& 7422	&       &       & 7420  & 7486        & 7461        & 7474  \\
3    & 4s$^2$4p$^4$($^3$P)4d $^4$D     & 1.5 &37&       &       &       & 12199 & 12178       & 12357       & 12165 \\
4    & 4s$^2$4p$^4$($^3$P)4d $^4$P     & 0.5 &45&	&       &       & 12370 & {\bf 12377} & {\bf 12545} & 12347 \\
5    & 4s$^2$4p$^4$($^3$P)4d $^4$D     & 2.5 &44& 12322 & 12322 &       & 12379 & {\bf 12362} & {\bf 12540} & 12353 \\
6    & 4s$^2$4p$^4$($^3$P)4d $^4$F     & 3.5 &32& 12520 &       &       & 12603 & 12577       & 12745       & 12572 \\
7    & 4s$^2$4p$^4$($^1$S)4d $^2$D     & 1.5 &54&	&       &       & 13255 & 13266       & 13337       & 13221 \\
8    & 4s$^2$4p$^4$($^3$P)4d $^4$D     & 3.5 &41& 13754	&       &       & 13749 & 13816       & 13992       & 13810 \\
9    & 4s$^2$4p$^4$($^3$P)4d $^2$P     & 0.5 &34&       &       &       & 13758 & 13886       & 14066       & 13855 \\
10   & 4s$^2$4p$^4$($^3$P)4d $^4$F     & 4.5 &68& 13811 &       &       & 13850 & 13893       & 14063       & 13916 \\
11   & 4s$^2$4p$^4$($^1$S)4d $^2$D     & 2.5 &45&	&       &       & 14889 & 15006       & 15083       & 14948 \\
12   & 4s$^2$4p$^4$($^3$P)4d $^4$P     & 1.5 &24& 15231 & 15231 & 15201 & 15262 & 15406       & 15468       & 15287 \\
13   & 4s$^2$4p$^4$($^3$P)4d $^2$D     & 2.5 &22& 15465 & 15465 & 15446 & 15513 & 15655       & 15723       & 15524 \\
14   & 4s4p$^6$              $^2$S     & 0.5 &67& 16380 & 16380 &       & 16268 & 16560       & 16867       & 16470 \\
15   & 4s$^2$4p$^4$($^3$P)4d $^4$D     & 0.5 &78&	&       &       & 19073 &             & 19395       & 19203 \\
16   & 4s$^2$4p$^4$($^3$P)4d $^4$D$_a$ & 1.5 &34&	&       &       & 19458 &             & 19790       & 19598 \\
17   & 4s$^2$4p$^4$($^3$P)4d $^4$F     & 2.5 &45&	& 19739 &       & 19809 & 19996       & 20126       & 19952 \\
18   & 4s$^2$4p$^4$($^1$D)4d $^2$G     & 3.5 &59&	&       &       & 19900 &             & 20203       & 20040 \\
19   & 4s$^2$4p$^4$($^3$P)4d $^4$D$_a$ & 3.5 &42&	&       &       & 20814 &             & 21221       & 21049 \\
20   & 4s$^2$4p$^4$($^1$D)4d $^2$P     & 1.5 &47&	&       &       & 21080 & 21406       & 21510       & 21309 \\
21   & 4s$^2$4p$^4$($^1$D)4d $^2$S     & 0.5 &42& 21355 & 21355 &       & 21196 & {\bf 21707} & {\bf 21759} & 21418 \\
22   & 4s$^2$4p$^4$($^3$P)4d $^2$D$_a$ & 2.5 &31& 21355 & 21355 & 21363 & 21239 & {\bf 21599} & {\bf 21678} & 21433 \\
23   & 4s$^2$4p$^4$($^3$P)4d $^2$P     & 1.5 &43& 21355 & 21355 & 21363 & 21261 & {\bf 21654} & {\bf 21683} & 21467 \\
24   & 4s$^2$4p$^4$($^3$P)4d $^2$F     & 2.5 &43&       &       &       & 21294 & {\bf 21651} & {\bf 21745} & 21540 \\
25   & 4s$^2$4p$^4$($^1$D)4d $^2$G     & 4.5 &68&	&       &       & 21326 & {\bf 21597} & {\bf 21726} & 21570 \\
26   & 4s$^2$4p$^4$($^1$D)4d $^2$D     & 2.5 &29&	& 21761 &       & 21567 & 21897       & 22026       & 21803 \\
27   & 4s$^2$4p$^4$($^1$D)4d $^2$F     & 3.5 &56&	&       &       & 21887 & 22221       & 22349       & 22136 \\
28   & 4s$^2$4p$^4$($^3$P)4d $^2$D     & 1.5 &39&	&       &       & 23041 & 23462       & 23475       & 23240 \\
29   & 4s$^2$4p$^4$($^3$P)4d $^2$P$_a$ & 0.5 &36&	&       &       & 23498 & 24025       & 23947       & 23701 \\
30   & 4s$^2$4p$^4$($^1$S)4d $^2$D$_a$ & 1.5 &34&	&       &       & 28316 &             & 28844       & 28682 \\
31   & 4s$^2$4p$^4$($^1$S)4d $^2$D$_a$ & 2.5 &37&	&       &       & 28939 &             & 29515       & 29365 \\
\ns
\mr
     &  $MSD_{[29]}$                   &     &  &       &       &       & 73    & 164         & 276         & 75    \\
     &  $MSD_{[22]}$                   &     &  &       &       &       & 103   & 202         & 298         & 97    \\
     &  $MSD_{[25]}$                   &     &  &       &       &       & 82    & 206         & 259         & 73    \\
\ns
\br
\endTable



The calculated level energies are presented in \tref{t1}. 
The level indices, total angular momenta $J$, configurations and $LS$ terms are 
also given in \tref{t1}. Presented percentage contributions of the $LS$ terms are 
further applied to make level assignment. The term identification is rather 
rough and formal; it is executed according to the largest weight in the CI 
wavefunction expansion. Therefore, for some levels with the same total angular 
momentum $J$, the same $LS$-term is attributed. In these instances, the $LS$ 
term has additional index $"a"$. The duplication of term assignments is not 
unusual for the multicharged ions as the $LS$-coupling is not good enough for 
highly-charged ions. 

Along with our QR results, we present the data from the NIST database \cite{29}
which are the same as in \cite{04}. The experimental level energies determined
from the experimental transition wavelengths given in \cite{22,25} are also
presented. As we have mentioned in \sref{intro}, one can find results of
different accuracy for the W$^{39+}$ ion. In order to make \tref{t1} as concise
as possible, we include the theoretical results only from \cite{23,33,35}.
Level energies in \cite{23} were determined using the relativistic parametric 
potential code RELAC. Unfortunately, the list of presented energy levels is
not complete for the configurations considered here. The level energies from
\cite{33} almost completely correspond to the GRASP3 results from \cite{32}.
The GK$_{\mathrm{CV}}$ results are taken from \cite{35}. These data
are produced using the largest CI expansion and are the closest to the 
experimental level energies. 

All the levels are presented according to the
QR energy increasing order. Such an ordering completely agrees with that of 
GK$_{\mathrm{CV}}$ results from \cite{35} and also corresponds to the order
of the experimental data. We must underline that, for the group of three levels
with the same calculated energies, we assign the level numbers 21, 22, and 23.
The same assignment have been done in \cite{35}, whereas the level $J=2.5$
from that group was given the level number 25 in \cite{32}. As it is highlighted
in \cite{35}, the ordering of some energy levels given in \cite{23,33} does not
correspond to the ordering following from the most accurate results in \cite{35}.
Consequently, it does not correspond to  the QR results. We mark those levels
in the bold case in \tref{t1}. At the end of this table, we present the
mean-square-deviations ($MSD$) determined by formula:

\begin{equation}
\fl
MSD = \left(\frac{\sum_N (E_N^{\mathrm{th}} - E_N^{\mathrm{exp}})^2(2J_N+1)}
{\sum_N (2J_N+1)} \right)^{1/2}.
\label{msd}
\end{equation}

Since several works give different sets of energy levels and slightly different 
energy values, we determine the $MSD$ for each set of the experimental level 
energies. These $MSD$ are distinguished by different indices. As one can clearly 
see, the QR results are more accurate compared to those from \cite{22} and
\cite{33}; their accuracy is almost the same as that of GK$_{\mathrm{CV}}$ 
results from \cite{35}. We must admit, that such an accuracy is achieved by
applying significantly large CI wavefunction expansion compared to the expansion
adopted in \cite{35}. This is caused by the fact that our calculations employ
the basis of the quasirelativistic RO, whereas the Dirac-Fock equation solutions 
are used in \cite{35} and other mentioned theoretical studies. Moreover, the 
same RO basis is employed both for even-parity configurations and for those
of odd parity. Therefore an additional correlation inclusion is necessary.
So the CI expansion where the CSFs with the same parity and principal quantum 
numbers are included (GK$_2$ in \cite{35}) produces $MSD_{[29]} = 9400$\,cm$^{-1}$ 
whereas our QR calculation with the same expansion produces only 
$MSD_{[29]} = 35900$\,cm$^{-1}$.

A very nice agreement is evident when we compare our QR percentage contributions
from \tref{t1} to analogous data from the table 2 in \cite{35}. Almost all 
contributions agree within $1\%$, only for the levels 14, 20, 23, and 26 the 
deviations reach $2\%$. Usually the QR percentage contributions are slightly
lower compared to those of GK$_{\mathrm{CV}}$ from \cite{35}. This happens 
because the larger CI wavefunction expansion is employed in our calculation.

In general, this kind of agreement of the QR and GK$_{\mathrm{CV}}$ results is
very encouraging as calculations are performed using basically different 
approximations. Likewise in QR results, GK$_{\mathrm{CV}}$ calculations produce
above-mentioned levels with the same main percentage contributions in the $LS$
coupling and their level assignement matches that of our QR data. 
When we compare the QR percentage contributions to those from \cite{33}, their
agreement is nice also for most levels. The differences of $1 - 2\%$ can be
observed not only for the largest (main) contributions but also for the other
most important components of eigenfunctions. We do not present these 
contributions in the present work. The large deviations up to $6\%$ between
the QR results and the data from \cite{33} are observed for the levels 
$20 - 23$, and for 26. Furthermore, similar large deviations of the percentage
contributions for these levels appear when the data from \cite{33} and \cite{35} 
are compared. It must be underlined, that energy ordering for these levels given 
in \cite{33} does not correspond to that of QR or GK$_{\mathrm{CV}}$ calculations.


\Table{\label{t2}
The radiative lifetimes (in ns) of the W$^{39+}$ levels.
}
\br
\multicolumn{1}{l}{$N$}&
\multicolumn{1}{c}{$J$}&
\multicolumn{1}{c}{QR}&
\multicolumn{1}{c}{MCDF[33]}\\
\mr
2  & 0.5 & 1.30E$+$2 & 1.31E$+$2       \\
3  & 1.5 & 3.97E$-$1 & 3.22E$-$1       \\
4  & 0.5 & 8.79E$-$2 & 7.15E$-$2       \\
5  & 2.5 & 7.85E$-$1 & 6.34E$-$1       \\
6  & 3.5 & 3.70E$+$6 & {\bf 4.76E$+$6} \\
7  & 1.5 & 4.71E$+$0 & {\bf 2.68E$+$0} \\
8  & 3.5 & 2.66E$+$4 & 2.24E$+$4       \\
9  & 0.5 & 2.00E$-$1 & {\bf 1.04E$-$1} \\
10 & 4.5 & 4.37E$+$4 & 3.70E$+$4       \\
11 & 2.5 & 8.40E$+$2 & {\bf 3.06E$-$1} \\
12 & 1.5 & 2.56E$-$3 & 2.36E$-$3       \\
13 & 2.5 & 1.86E$-$3 & 1.75E$-$3       \\
14 & 0.5 & 1.62E$-$3 & 1.66E$-$3       \\
15 & 0.5 & 1.19E$-$1 & 1.22E$-$1       \\
16 & 1.5 & 5.63E$-$1 & 5.11E$-$1       \\
17 & 2.5 & 2.36E$-$2 & 2.07E$-$2       \\
18 & 3.5 & 2.72E$+$2 & 2.60E$+$2       \\
19 & 3.5 & 1.45E$+$2 & 1.36E$+$2       \\
20 & 1.5 & 2.83E$-$3 & {\bf 1.32E$-$3} \\
21 & 0.5 & 4.19E$-$4 & 3.71E$-$4       \\
22 & 2.5 & 5.42E$-$4 & 5.21E$-$4       \\
23 & 1.5 & 7.29E$-$4 & {\bf 1.00E$-$3} \\
24 & 2.5 & 2.23E$+$1 & {\bf 1.73E$-$2} \\
25 & 4.5 & 2.58E$+$2 & 2.45E$+$2       \\
26 & 2.5 & 9.42E$-$3 & {\bf 1.66E$-$2} \\
27 & 3.5 & 2.18E$+$2 & 2.05E$+$2       \\
28 & 1.5 & 1.33E$-$3 & 1.22E$-$3       \\
29 & 0.5 & 8.16E$-$4 & 7.44E$-$4       \\
30 & 1.5 & 6.55E$-$4 & 6.22E$-$4       \\
31 & 2.5 & 7.53E$-$1 & {\bf 4.37E$-$1} \\
\br
\endTable


To determine the spectroscopic parameters in the QR approach, we have
calculated the E1, M2 and E3 radiative transition data for the transitions
between the levels of different-parity configurations. The M1 and E2 radiative 
transitions have been calculated between the levels of the same-parity
configurations. Calculated radiative transition data are used to determine
the lifetimes $\tau$ of the excited levels. These data are available from
\tref{t2} where the level indices correspond to those in \tref{t1}.
The total angular momenta $J$ of the corresponding levels are also given.

In this table we compare the results of our quasirelativistic calculations
with the data from \cite{33}. The agreement is really nice for most levels.
The values $\tau_{[33]}$ which differ from the $\tau_{\mathrm{QR}}$ values by
more than $20\%$ are marked in bold face. In most cases, it happens for the
values where the radiative lifetimes are relatively large. As in the case of the
energy spectra, the radiative lifetimes from \cite{33} agree very well with
those from \cite{32} determined in the GRASP3 approximation. One can notice
only one substantial difference. The levels 22 and 24 with $J=2.5$ are swapped
around in \cite{33} compared to their positions in \cite{32}. This change 
makes the agreement with the QR results significantly better. Nevertheless,
substantial differences between our QR data and those from 
\cite{33} still exists for the levels 11 and 24.

One more point has to be underlined when one discusses the radiative lifetimes.
As it is already demonstrated in \cite{33}, there exists number of levels 
arising from the odd-parity configurations 4s$^2$4p$^3$4d$^2$, 4s4p$^5$, and 
4s$^2$4p$^4$ which are located below two high-lying even-parity levels 30 and 31.
The radiative transitions to those odd-parity levels were not calculated in the
present work. They were not presented in \cite{33} also. We have to mention that
the transition energies for those transitions are relatively small therefore
corresponding transition probabilities can not be large. Nevertheless, determined 
radiative lifetimes $\tau_{\mathrm{QR}}$ are probably slightly overestimated.


\setlength{\tabcolsep}{4pt}
\renewcommand{\arraystretch}{0.52}
\Table{\label{t3}
The W$^{39+}$ emission transition probabilities $A$ (in s$^{-1}$) 
determined in QR and several relativistic approximations and the
percentage deviations $k$ (see Eq.\eref{dev}).
}
\br
\multicolumn{1}{c}{}&
\multicolumn{1}{c}{}&
\multicolumn{1}{c}{}&
\multicolumn{1}{c}{QR}&
\multicolumn{1}{c}{MCDF}&
\multicolumn{1}{c}{MCDF}&
\multicolumn{1}{c}{MCDF}&
\multicolumn{1}{c}{}\\
\multicolumn{1}{c}{}&
\multicolumn{1}{c}{}&
\multicolumn{1}{c}{}&
\multicolumn{1}{c}{}&
\multicolumn{1}{c}{GK$_2$[35]}&
\multicolumn{1}{c}{GK$_{\mathrm{CV}}$[35]}&
\multicolumn{1}{c}{GRASP[33]}&
\multicolumn{1}{c}{$k$}\\
\mr
2   & 1   & M1 & 7.33(06) &          &           & 7.29(06) & -1\%     \\
2   & 1   & E2 & 3.40(05) &          &           & 3.28(05) & -4\%     \\
3   & 1   & E1 & 2.52(09) &          &           & 3.10(09) & 23\%     \\
4   & 1   & E1 & 1.12(10) &          &           & 1.38(10) & 23\%     \\
5   & 1   & E1 & 1.27(09) & 1.51(09) & 1.42(09)  & 1.58(09) & 24\%     \\
6   & 1   & M2 & 1.51(02) & 	        &           & 1.41(02) & -7\%     \\
6   & 5   & M1 & 9.06(01) & 	        &           & 6.89(01) & -24\%    \\
6   & 1   & E3 & 2.84(01) & 	        &           & 	        &  	       \\
7   & 1   & E1 & 2.03(08) & 	        &           & 3.61(08) & 78\%     \\
7   & 2   & E1 & 9.65(06) & 	        &           & 1.24(07) & 29\%     \\
8   & 5   & M1 & 2.04(04) & 	        &           & 2.46(04) & 21\%     \\
8   & 1   & M2 & 1.06(04) & 	        &           & 1.18(04) & 11\%     \\
8   & 6   & M1 & 6.63(03) & 	        &           & 8.38(03) & 26\%     \\
9   & 1   & E1 & 3.41(09) & 	        &           & 8.44(09) & 148\%    \\
9   & 2   & E1 & 1.59(09) & 	        &           & 1.18(09) & -26\%    \\
10  & 6   & M1 & 2.29(04) & 	        &           & 2.70(04) & 18\%     \\
11  & 1   & E1 & 1.12(06) & 	        &           & 3.27(09) & 2.9$\cdot 10^5$\% \\
11  & 7   & M1 & 4.28(04) & 	        &           & 5.60(04) & 31\%     \\
12  & 1   & E1 & 3.90(11) & 4.10(11) & 3.99(11)  & 4.24(11) & 9\%      \\
13  & 1   & E1 & 5.39(11) & 5.67(11) & 5.53(11)  & 5.71(11) & 6\%      \\
14  & 1   & E1 & 5.91(11) & 6.17(11) & 6.03(11)  & 5.78(11) & -2\%     \\
14  & 2   & E1 & 2.65(10) & 	        &           & 2.55(10) & -3\%     \\
15  & 1   & E1 & 6.57(09) & 	        &           & 6.47(09) & -2\%     \\
15  & 2   & E1 & 1.84(09) & 	        &           & 1.72(09) & -6\%     \\
16  & 2   & E1 & 1.66(09) & 	        &           & 1.60(09) & -4\%     \\
16  & 1   & E1 & 1.14(08) & 	        &           & 3.53(08) & 210\%    \\
17  & 1   & E1 & 4.24(10) & 3.99(10) & 4.03(10)  & 4.84(10) & 14\%     \\
18  & 6   & M1 & 2.87(06) & 	        &           & 3.00(06) & 5\%      \\
18  & 5   & M1 & 4.66(05) & 	        &           & 4.81(05) & 3\%      \\
18  & 6   & E2 & 1.26(05) & 	        &           & 1.34(05) & 7\%      \\
19  & 10  & M1 & 2.77(06) & 	        &           & 2.95(06) & 7\%      \\
19  & 8   & M1 & 1.99(06) & 	        &           & 2.08(06) & 4\%      \\
19  & 13  & M1 & 1.29(06) & 	        &           & 1.28(06) & -1\%     \\
19  & 11  & M1 & 4.97(05) & 	        &           & 7.02(05) & 41\%     \\
19  & 5   & M1 & 1.28(05) & 	        &           & 1.35(05) & 6\%      \\
19  & 10  & E2 & 9.66(04) & 	        &           & 1.06(05) & 10\%     \\
20  & 1   & E1 & 3.54(11) & 3.57(11) & 5.18(11)  & 7.60(11) & 115\%    \\
21  & 1   & E1 & 2.38(12) & 2.49(12) & 2.42(12)  & 2.69(12) & 13\%     \\
22  & 1   & E1 & 1.85(12) & 1.97(12) & 1.86(12)  & 1.92(12) & 4\%      \\
23  & 1   & E1 & 1.37(12) & 1.46(12) & 1.18(12)  & 9.91(11) & -28\%    \\
24  & 1   & E1 & 3.79(07) & 	        &           & 5.78(10) & 1.5$\cdot 10^5$\% \\
24  & 8   & M1 & 4.56(06) & 	        &           & 4.80(06) & 5\%      \\
24  & 11  & M1 & 2.09(06) & 	        &           & 2.10(06) & 0\%      \\
25  & 10  & M1 & 2.91(06) & 	        &           & 3.06(06) & 5\%      \\
25  & 8   & M1 & 6.68(05) & 	        &           & 7.05(05) & 5\%      \\
25  & 10  & E2 & 1.25(05) & 	        &           & 1.36(05) & 8\%      \\
25  & 8   & E2 & 8.56(04) & 	        &           & 3.39(04) & -60\%    \\
26  & 1   & E1 & 1.06(11) & 1.13(11) & 7.78(10)  & 6.01(10) & -43\%    \\
27  & 10  & M1 & 2.55(06) & 	        &           & 2.67(06) & 5\%      \\
27  & 8   & M1 & 8.18(05) &          &           & 8.80(05) & 8\%      \\
27  & 11  & M1 & 6.49(05) &          &           & 6.24(05) & -4\%     \\
27  & 13  & M1 & 2.21(05) &          &           & 3.12(05) & 41\%     \\
27  & 11  & E2 & 9.49(04) &          &           & 1.04(05) & 10\%     \\
27  & 10  & E2 & 8.62(04) &          &           & 9.85(04) & 14\%     \\
27  & 17  & M1 & 5.30(04) &          &           & 6.42(04) & 21\%     \\
28  & 2   & E1 & 6.59(11) &          &           & 6.98(11) & 6\%      \\
28  & 1   & E1 & 9.41(10) &          &           & 1.19(11) & 26\%     \\
29  & 2   & E1 & 1.21(12) &          &           & 1.34(12) & 10\%     \\
29  & 1   & E1 & 1.14(10) &          &           & 1.62(09) & -86\%    \\
30  & 2   & E1 & 1.53(12) &          &           & 1.61(12) & 5\%      \\
31  & 1   & E1 & 1.31(09) &          &           & 2.27(09) & 73\%     \\
\br
\endTable


In \tref{t3} we present the radiative transition parameters for all excited
levels of the W$^{39+}$ ion configurations 4s$^2$4p$^5$, 4s$^2$4p$^4$4d and 
4s4p$^6$. We present only those radiative transitions probabilities $A$ which 
play an important role in determining the level radiative lifetimes $\tau$.

The initial and the final level indices from \tref{t1} and the radiative 
transition type are given to describe a particular transition. We compare our 
determined transition probabilities A$_{\mathrm{QR}}$ with the data from 
\cite{35} calculated in the GK$_2$ and GK$_{\mathrm{CV}}$ approximations and
with the transition probabilities from \cite{33}. 

It is worth mentioning that 
inclusion of the 3d-shell polarization, which corresponds transition from the
approximation GK$_2$ to the GK$_{\mathrm{CV}}$ in \cite{35}, makes the 
agreement of QR results and the latter data better for most transitions. 

The transition probability values in the GK$_{\mathrm{CV}}$ approximation agree
with our QR data within few percent limits for more than half of considered
lines. The main exclusions from that path are the radiative transitions from the 
levels 20 and 26 to the lowest level 1. In this case, the inclusion of the 
3d-shell polarization significantly changes calculated transition probability 
values. Such a behavior is not 
observed for the remaining transitions. Transition probabilities from the levels 
20 and 26 determined in the GK$_{\mathrm{CV}}$ approximation are the only ones
from \cite{35} which differ by more than $20\%$ from our QR results.

The comparison of the transition probabilities from \cite{33} with our data
gives a different picture. Their percentage deviation $k$

\begin{equation}
k=\left( \frac{A_{[33]}}{A_{\mathrm{QR}}}-1 \right) \cdot 100\%
\label{dev}
\end{equation}
are presented in the last column of \tref{t3}. One can see from that table that
almost a half of transitions of various types agree within $10\%$. There is a
group of 21 transitions where deviations reach up to $50\%$. For 9 transitions,
the deviations are over $50\%$. Two E1 transition probabilities from the levels 
11 and 24 to the level 1 are quite exceptional because their values differ by
three orders of magnitude. Unfortunately these transition probabilities are not
given in \cite{35}. It is very difficult to find a reason of such an essential
difference. As we have explained before, the main $LS$ contributions in the CI 
wavefunction expansions for these levels agree nicely. Furthermore, there is 
quite good agreement of the transition probabilities from the levels 11 and 24
to other lower levels. There is one more point worth mentioning here. If one 
checks the $gf$ values for these particular transitions $11-1$ and $24-1$ given 
in \cite{32}, it is evident that they differ by more than one order of magnitude
going from one approximation to another. Therefore we conclude that it is 
necessary to perform an independent calculation in order to find correct answer.
 
The total M1 and E2 transition probability value 
$A_{\mathrm{M1}}+A_{\mathrm{E2}}= 7.65 \times 10^6$ s$^{-1}$ is given in 
\cite{10}. This value agrees very well with our QR value 
$A_{\mathrm{M1}}+A_{\mathrm{E2}} = 7.67 \times 10^6$ s$^{-1}$.

In the present work we also have determined parameters for the E3 transition 
from the excited configuration levels to the ground configuration. As it has 
been demonstrated in \cite{16, 17}, this type of the radiative transitions 
can significantly affect the calculated radiative lifetime $\tau$ values of 
the metastable levels for the tungsten and other ions with the open 4d shell. 
This feature occur for the level 6 of the investigated W$^{39+}$ ion. 
When we include the E3 transition from the level 6 to the level 1, it changes
the determined lifetime $\tau$ value by $10\%$. That indicates that such a 
transition can affect calculated branching ratios for the decay of this level.

There exists another level, the level 8, where the 
M2 transition to the ground configuration influences the calculated radiative 
lifetime $\tau$ value. For all other levels of the excited configurations 
4s$^2$4p$^4$4d and 4s4p$^6$, the radiative lifetimes are well-determined by the 
E1, M1 and E2 transitions. We have no doubt that calculation of the M2 and E3 
radiative transitions to the ground configurations is also important for the 
excited configuration levels of the ions with the open 4p$^N$ shell.

\section{Summary and conclusions}
\label{summ}

The developed quasirelativistic approach facilitate multiconfiguration
calculations employing the broad CI wavefunction basis of the admixed 
configurations. For the inclusion of the correlation effects, the transformed 
radial orbitals are employed. The effectiveness of this approach have been 
proved in many previous calculations. We include almost 1000 configurations in 
the CI wavefunction expansion for the excited configurations with the reduced 
number of CSFs exceeding 400\,000. 

The admixed configurations are generated by the virtual excitation of electrons 
from the external $4l$ and from the inner $3l$ shells. This leads to the 
efficient inclusion of the correlation effects. As a result, the calculated 
level energies for the excited configurations 4s$^2$4p$^4$4d and 4s4p$^6$ agree 
very well with the experimental data. 

Determined parameters of the radiative E1 transitions agree well with the most 
reliable relativistic calculations. That makes us to believe that the transition
probability values for other considered transitions are
accurate enough. Produced QR data make a complete set for three investigated
configurations. Hence they can be applied both for the interpretation of new
experimental results and for modeling spectra of high-temperature plasma.

We have demonstrated that the E3 radiative transitions along with the M2 
transitions to the ground configuration 4s$^2$4p$^5$ must be included in 
order to determine the correct radiative lifetime values for the metastable 
levels of the excited configurations 4s$^2$4p$^4$4d and 4s4p$^6$. 
This conclusion bears similarity with the outcome from our investigation of 
the multicharged ions with the open 4d shell.

\ack
The current research is funded by the European Social Fund under the Global 
Grant measure, Project No. VP1-3.1-{\v S}MM-07-K-02-013.

\end{document}